\documentclass[aps,prd,preprint,groupedaddress,nofootinbib,showpacs,eqsecnum]{revtex4}
\usepackage[T1]{fontenc}
\usepackage{graphicx,epsf,color,amsmath}
\usepackage[caption=false]{subfig}
\usepackage{booktabs,multirow}

\def\sect#1{Sect.~{\ref{#1}}}

\def\fig#1{Fig.~{\ref{#1}}}

\def\tab#1{Table~{\ref{#1}}}

\def\eqn#1{Eq.~(\ref{#1})}
\def\eqns#1#2{Eqs.~(\ref{#1}) and (\ref{#2})}
\def\Eqn#1{Eq.~(\ref{#1})}

\def\NeqOne{{\mathcal{N}=1}}
\def\NeqTwo{{\mathcal{N}=2}}
\def\NeqFour{{\mathcal{N}=4}}
\def\eps{\epsilon}
\def\<{\langle}
\def\>{\rangle}
\def\pol{\varepsilon}
\def\tree{{\rm tree}}
\def\oneloop{{\rm 1\hbox{-}loop}}

\def\msym{{\NeqFour}}
\def\hmsg{\NeqFour,{\rm SG}}

\def\spa#1.#2{\left\langle#1\,#2\right\rangle}
\def\spb#1.#2{\left[#1\,#2\right]}
\def\Ord{{\cal O}}

\def\spa#1.#2{\left\langle#1\,#2\right\rangle}
\def\spb#1.#2{\left[#1\,#2\right]}
\def\ev#1{\textrm{ev}#1}
\def\mhv#1{\textsc{mhv}#1}

\def\statesum{\odot}
\def\tcdot{\!\cdot\!}
\def\hel#1#2#3#4{({\mskip-1mu}#1{\mskip-1mu}#2{\mskip-1mu}#3{\mskip-1mu}#4{\mskip-1mu})}

\def\Ff#1{F^{4}_{#1}}
\def\Ft#1{(F^{2}_{#1})^2}

\def\inputdeferred#1{}

\begin{document}
\hfuzz=15 pt

\title{Curvature-Squared Multiplets, Evanescent Effects\\
 and the U(1) Anomaly in $\NeqFour$ Supergravity}

\author{Zvi~Bern${}^{ab}$, Alex~Edison${}^a$, David~Kosower${}^{cd}$ and
             Julio~Parra-Martinez${}^{ab}$ \\$\null$\\}

\affiliation{
  ${}^a$Mani L. Bhaumik Institute for Theoretical Physics,
  Department of Physics and Astronomy,
  University of California at Los Angeles, Los Angeles, CA 90095, USA\\
  \vskip -4 mm 
  ${}^b$Kavli Institute for Theoretical Physics,
   \hbox{University of California, Santa Barbara, CA 93106, USA}\\
  \vskip -4 mm 
${}^c$\hbox{School of Natural Sciences, Institute for Advanced Study},\\
\hbox{1 Einstein Drive, Princeton, New Jersey 08540}\\
%
  \vskip -4mm 
 ${}^d$\hbox{Institut de Physique Th\'eorique, CEA, CNRS, Universit\'e Paris--Saclay},
  F--91191 Gif-sur-Yvette cedex, France \\
}

\begin{abstract}
	
We evaluate one-loop amplitudes of $\NeqFour$ supergravity in $D$
dimensions using the double-copy procedure that expresses gravity
integrands in terms of corresponding ones in Yang--Mills theory.  We
organize the calculation in terms of a set of gauge-invariant tensors,
allowing us to identify evanescent contributions. Among the latter, we find
the matrix elements of supersymmetric completions of curvature-squared
operators.  In addition, we find that such evanescent terms and the
$\text{U}(1)$-anomalous contributions to one-loop $\NeqFour$ amplitudes are
tightly intertwined.  The appearance of evanescent operators in
$\NeqFour$ supergravity and their relation to anomalies raises the
question of their effect on the known four-loop divergence in this
theory.  We provide bases of gauge-invariant tensors and corresponding projectors
useful for Yang--Mills theories as a by-product of our
analysis.
\end{abstract}

\pacs{04.65.+e, 11.15.Bt, 11.25.Db, 12.60.Jv \hspace{1cm}}

\maketitle

\section{Introduction and Review}
\label{sec:intro}

Recent explicit calculations have shown that gravity theories still
have perturbative secrets waiting to be revealed.  We have learned a
number of surprising lessons from these calculations: results in gravity theories can be
obtained directly from their Yang--Mills counterparts via a
double-copy
procedure~\cite{BCJLoop,DoubleCopy,Simplifying,ClassicalSolutions}; of
a curious disconnect between the leading two-loop divergence of
graviton amplitudes~\cite{GoroffSagnotti,vandeVen} and the
corresponding renormalization-scale
dependence~\cite{GBPureGravity,GBPureGravitySimple}; and about the
surprisingly tame ultraviolet behavior of certain supergravity
theories~\cite{N4GravThreeLoops,N5GravFourLoops,HalfMaxD5,KellyAttempt}.
These lessons augur more surprises to come.  In this paper we 
investigate the role of evanescent effects in the one-loop four-point
amplitude of $\NeqFour$ supergravity, along with its relation to the
$\text{U}(1)$ anomaly in the duality symmetry of this
theory~\cite{MarcusAnomaly, CarrascoAnomaly, RenataAnomaly}.

Evanescent effects arise from operators whose matrix elements vanish
when working strictly in four dimensions, but give rise to
nonvanishing contributions in dimensional regularization.  Such
contributions originate from the cancellation of poles against small
deviations in the four-dimensional limit; that is, they are due to
$\eps/\eps$ effects, where $\eps = (4-D)/2$ is the dimensional
regulator. Although such effects might at first appear to be a mere
technicality, they turn out to play an important
role~\cite{GBPureGravity} in understanding ultraviolet divergences of
Einstein gravity in the context of dimensional
regularization~\cite{GoroffSagnotti,vandeVen}.  In particular, the
Gauss--Bonnet operator is evanescent and appears as a one-loop
counterterm whose insertion at two loops contaminates the ultraviolet
divergence, but results in no physical consequences in the
renormalized amplitude.  An important question therefore is whether a
supersymmetric version of the Gauss-Bonnet operator appears in the
matrix elements of $\NeqFour$ supergravity. If such an operator exists
it would be important to determine its effects on the known four-loop
divergence~\cite{FourLoopN4Sugra} of the theory.

On the other hand, the $\NeqFour$ supergravity theory has an anomaly
in its $\text{U}(1)$ duality symmetry~\cite{MarcusAnomaly}. The
anomaly manifests itself in the failure of certain helicity amplitudes
which vanish at tree level to persist in vanishing at loop level.  In
the context of dimensional regularization these anomalous amplitudes
arise from $\eps/\eps$ effects, in much the same way as the usual
chiral anomaly arises in the 't Hooft--Veltman scheme~\cite{HVScheme}.
Refs.~\cite{CarrascoAnomaly,FourLoopN4Sugra} have suggested that the
$\text{U}(1)$ duality anomaly plays a key role in the four-loop
divergence of the theory~\cite{CarrascoAnomaly,FourLoopN4Sugra},
although a detailed explanation is still lacking.  In contrast to the
anomaly terms, it is unlikely that evanescent effects can alter any
physical quantity derived from scattering
amplitudes~\cite{GBPureGravity,GBPureGravitySimple}.  Nevertheless,
one may wonder if there any connections between the two phenomena,
given that both arise from $\eps/\eps$ effects.

In order to investigate these questions
we compute the one-loop
four-point amplitude of $\NeqFour$ supergravity in arbitrary
dimensions, using the double-copy procedure based on the duality
between color and kinematics~\cite{BCJ,BCJLoop}. The corresponding
helicity amplitudes were previously calculated using various
methods~\cite{Dunbar1995,Dunbar2011,OneLoopN4}.  Here, we use formal
polarizations in order to study evanescent effects, which are hidden
when four-dimensional helicity states are used.  The conclusion of our
study is two-fold: an evanescent contribution of the Gauss--Bonnet
type does appear in the pure-graviton amplitude of
$\NeqFour$ supergravity; and its effects are indeed intertwined with
the $\text{U}(1)$ duality anomaly.

We argue that the main evanescent contributions to the amplitude
correspond to the supersymmetric generalization of the
curvature-squared terms.  Off-shell forms of curvature-squared
operators are known for $\NeqOne$ and $\NeqTwo$
supergravity~\cite{FerraraGB,ButterN2GB}; but explicit forms 
of a supersymmetric extension of the Gauss-Bonnet 
curvature-squared operator are not known off shell in $\NeqFour$ supergravity.\footnote{Curvature-squared 
operators have been studied in the context of conformal supergravity~\cite{BergshoeffN4}.}
Nonetheless their matrix
elements can be computed directly using standard amplitude methods,
even without knowing their off-shell forms.  In contrast to the
nonsupersymmetric case, the coefficients of these matrix elements are
finite.  This turns out to be a consequence of the same $\eps/\eps$
cancellation that generates the anomaly.  As we will see, in the
context of the double-copy construction there is a single object that has
matrix elements that contribute to both the anomaly and evanescent
curvature-squared terms.

The double-copy structure implies that we can write the one-loop
four-point amplitude of $\NeqFour$ supergravity in terms of
pure-Yang--Mills theory building blocks, up to an overall factor.  We
can therefore employ a set of gauge-invariant tensors written
in terms of formal gluon polarization vectors to carry out the
calculation.  We present the results in terms of linearized field
strengths, which is natural for connecting to operators in a
Lagrangian and making manifest on-shell gauge invariance.  In order to
explore the evanescent properties we also construct tensors with
definite four-dimensional helicity properties.  We provide the tensors
in a form natural for use in color-ordered Yang--Mills theory, as well
as in a fully crossing-symmetric form natural in $\NeqFour$
supergravity.  Similar gauge-invariant tensors have recently been discussed
by Boels and Medina~\cite{Boels}.

In the Appendix we give details of the gauge-invariant tensors and
describe the construction of projectors for determining the
coefficient of the tensors in a given amplitude.  These projectors and
tensors are useful not only for $\NeqFour$ supergravity but can be
applied to four-gluon amplitudes at any loop order in any Yang--Mills
theory, including quantum chromodynamics (QCD). Because of their more
general usefulness we attach a \textsl{Mathematica\/}
file~\cite{Ancillary} that includes the two sets of tensors with
different symmetry properties, alongside the corresponding projectors.

This paper is organized as follows.  In \sect{sec:amplitude} we give 
the construction of the four-loop four-point amplitude of $\NeqFour$ supergravity
and describe the gauge-invariant tensors in terms of which the amplitudes are
constructed.  In \sect{sec:maptogravity} we give the results 
for the one-loop supergravity amplitudes.  Then in \sect{sec:R2multiplets} we 
identify evanescent curvature-squared terms in the amplitude. We show 
the connection of these terms to the $\text{U}(1)$ anomaly in \sect{sec:anomaly}.
We give our conclusions in \sect{sec:conclusion}.  An appendix describing the
gauge-invariant tensors and projectors is included.

\section{Construction of the One-Loop Amplitude}
\label{sec:amplitude}

In this section we construct the one-loop four-point amplitude of
$\NeqFour$ supergravity.  Details of the gauge-invariant tensors used
for expressing the results are found in the appendix.

\subsection{Color-Kinematics Duality and the Double Copy}
\label{ssec:doublecopy}

We apply the double-copy construction of gravity amplitudes based on
the duality between color and kinematics~\cite{BCJ,BCJLoop}. This has
previously been discussed in some detail in Ref.~\cite{OneLoopN4} for 
the one-loop
amplitudes of $\NeqFour$ supergravity.  In contrast to the earlier
construction, we use $D$-dimensional external states instead of
four-dimensional ones, in order to have access to evanescent effects.

Amplitudes of half-maximal supergravity in $D$ dimensions can be
obtained through a double copy, where one factor is derived from
maximally supersymmetric Yang--Mills theory (MSYM), and the other from
pure Yang--Mills (YM) theory. In four dimensions, this gives us
amplitudes in $\NeqFour$ supergravity in terms of a product of
$\NeqFour$ and pure Yang--Mills theory.  Alternatively, one may also construct
$\NeqFour$ supergravity amplitudes using two copies of $\NeqTwo$
super-Yang--Mills (SYM) theory, as shown in Ref.~\cite{N2TimesN2}.
This latter construction is, however, more complicated, and
furthermore includes unwanted matter multiplets. We use the
simpler construction.

The double-copy construction starts from the integrands of two
Yang--Mills gauge-theory amplitudes, written in terms of purely cubic
diagrams.  In a Feynman-diagram language, four-point vertices can
always be ``blown up'' into a product of three-point vertices,
possibly with the exchange of a fictitious tensor field.  The
representation of one-loop amplitudes is, 
\begin{equation}
 {\cal A}^\oneloop_m =  {i} {g^{m}}
\int \frac{d^{D} p}{ (2 \pi)^{D}}\; \sum_{j\in\textrm{ICD}}  
  \frac{1}{S_j}  \frac {n_j c_j}{\prod_{\alpha_j}{p^2_{\alpha_j}}}\,,
\label{LoopGauge}
\end{equation}
where the sum runs over the independent cubic diagrams (ICD) labeled
by $j$, while the $c_j$ and $n_j$ are the color factors and kinematic
numerators associated with each diagram. The factor $1/S_j$ accounts
for the usual diagram symmetry factors and the product over $\alpha_j$
runs over the Feynman propagators $1/p_{\alpha_j}^2$ for diagram $j$.
If the kinematic numerators can be arranged to satisfy the same
algebraic properties as adjoint representation color factors, that is
so that Jacobi relations hold,
\begin{equation}
c_i + c_j + c_k = 0\;  \Rightarrow \;  n_i + n_j + n_k = 0 \,,
\label{BCJDuality}
\end{equation}
along with all anti-symmetry properties, then we can obtain gravity
integrands and thence amplitudes by replacing the color factors 
$c_j$ in \eqn{LoopGauge} by the second Yang--Mills theory's 
kinematic numerators,
\begin{equation}
c_i \rightarrow \tilde{n}_i\,.
\label{ColorSubstitutionRule}
\end{equation}
We do this while keeping the original kinematic factors $n_j$ of the
first Yang--Mills theory.  A similar procedure holds for 
particles in the fundamental representation~\cite{HenrikFundamental}.

\begin{figure}[tb]
\centering
\includegraphics[width=0.50\textwidth]{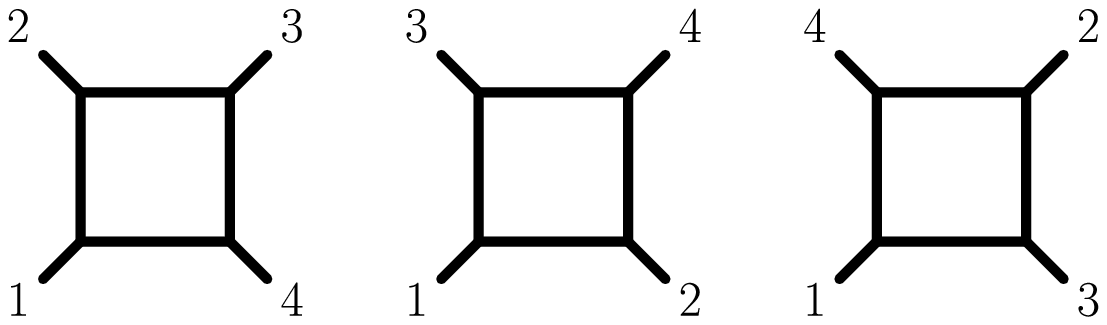}
\caption{Box diagrams of the one-loop four-point amplitude of
  $\NeqFour$ supergravity.}
\label{oneBox}
\end{figure}

The one-loop four-point amplitude of $\NeqFour$ supergravity is
easy to construct via the double-copy construction, because the
$\NeqFour$ MSYM numerators are especially simple~\cite{GSB}.  The
numerators of triangle and bubble diagrams vanish, and the box
integrals illustrated in \fig{oneBox} have kinematic numerators
proportional to the tree amplitude,
\begin{equation}
  n_{1234}= n_{1342}= n_{1423}= s\,t 
  A^\tree_\msym(1,2,3,4)\,,
\end{equation}
where we define the usual Mandelstam invariants,
\begin{equation}
s = (k_1 + k_2)^2\,, \hskip 1cm t = (k_2 + k_3)^2\,, \hskip 1cm 
u = (k_1 + k_3)^2\,.
\end{equation}
These numerators trivially satisfy the dual Jacobi identities
in \eqn{BCJDuality}.  Thus, the $\NeqFour$ supergravity one-loop
amplitude is
\begin{equation}
M_{\hmsg}^{\text{1-loop}}(1,2,3,4) =
i  s t A^{\tree}_{\msym}(1, 2, 3, 4)
\Bigl(I_{1234}[n_{1234,p}] + I_{1342}[n_{1342,p}] + I_{1423}[n_{1423,p}] \Bigr) \,,
\label{integrand}
\end{equation}
where we have stripped the gravitational coupling, and where
\begin{equation}
I_{1234}[n_{1234,p}] \equiv
\int \frac{d^D p}{(2\pi)^D} \frac{n_{1234,p}}{p^2 (p-k_1)^2 (p-k_1-k_2)^2
      (p+k_4)^2} \,,
\end{equation}
is the first box integral in \fig{oneBox} and ${n_{1234,p}}$ is the
pure Yang--Mills kinematic numerator given in Eq.~(3.5) of
Ref.~\cite{ColorKinOneTwoLoops}.   
We can restore the coupling to the supergravity amplitude via,
\begin{equation}
{\cal M}_{\hmsg}^{\tree}(1,2,3,4) = \Bigl(\frac{\kappa}{2} \Bigr)^2 M_{\hmsg}^{\tree}(1,2,3,4) \,, 
\end{equation}
at tree level, and
\begin{equation}
{\cal M}_{\hmsg}^{\oneloop}(1,2,3,4) = \Bigl(\frac{\kappa}{2} \Bigr)^4 M_{\hmsg}^{\oneloop}(1,2,3,4) \,, 
\end{equation}
at one loop.
The coupling is related to Newton's constant via $\kappa^2 =
32 \pi G_N$.  An alternate form of \eqn{integrand} is,
\begin{align}
M_{\hmsg}^{\text{1-loop}} (1,2,3,4) = &
i  s t A^{\tree}_\msym (1, 2, 3, 4) \nonumber\\
& \times \Bigl(A^\oneloop(1,2,3,4) + A^\oneloop(1,3,4,2) + A^\oneloop(1,4,2,3) \Bigr) \,,
\label{integrandA}
\end{align}
where $A^\oneloop(1,2,3,4)$ is the color-ordered one-loop amplitude of
pure Yang--Mills theory.  The difference between Eqs.~(\ref{integrand})
and~(\ref{integrandA}) cancels
in the permutation sum.  The second form makes gauge invariance
manifest, as the building blocks are gauge-invariant
color-ordered amplitudes.
We use the form in \eqn{integrand} to evaluate the
amplitude explicitly.


\subsection{Gauge-Invariant Building Blocks}
\label{ssec:tensors}

The relatively simple double-copy structure of the one-loop four-point
$\NeqFour$ supergravity amplitude displayed in~\eqn{integrandA}
makes manifest a factorization into the product of an MSYM tree amplitude
and a sum over the three distinct permutations of the one-loop
color-ordered amplitude of pure Yang--Mills theory.  This suggests
that we can obtain a convenient organization of the supergravity
amplitude by first decomposing the Yang--Mills amplitudes into
gauge-invariant contributions.  We do so using bases of local on-shell
`gauge-invariant tensors'.  By gauge-invariant tensors here we mean
polynomials in $(\pol_i\cdot\pol_j),(k_i\cdot\pol_j)$ and $(k_i\cdot
k_j)$ that vanish upon replacing $\pol_i$ by $k_i$. These tensors are
distinct only if they differ after imposing on-shell conditions.
We can build such tensors by starting with tree-level
four-point scattering amplitudes for external gluons, for example, or with
four-point matrix elements of local gluonic operators, and then multiplying
by appropriate factors of $s$, $t$, or $u$ to make the quantities
local.  Boels and Medina~\cite{Boels} have also recently constructed
such tensors.

In the Appendix we present two different bases.  In the first, we
impose definite cyclic symmetry; this yields a basis natural for
color-ordered Yang--Mills amplitudes.  In the second, we impose
definite symmetry under crossing, making them natural for
supergravity.  Associated with each gauge-invariant tensor is a
projector built out of momenta and conjugate polarization vectors.
When applied to an integrand, it yields the coefficient of the given
tensor.  Integrating the coefficient then yields the coefficient of
the tensor in the amplitude.  This type of projection to a basis of
gauge-invariant tensors has been used in Ref.~\cite{ProjectorMethod}.
We stress that the first of these bases is directly useful in
gauge-theory calculations.  We refer the reader to the Appendix for
more details about the bases, their properties, their construction and
the projection techniques.  We also make these tensors and projectors
available in a ancillary \textsl{Mathematica\/} file~\cite{Ancillary}.

We apply this projection technique to the integrand in \eqn
{integrand}. This reduces the numerators to sums of products of
inverse propagators and external kinematics.  The integrand is then
expressed as a sum over tensors, with each coefficient expressed in
terms of the scalar box and simpler triangle and bubble integrals
that are easy to evaluate (via Feynman parameterization, for example).
The scalar box integral is taken from
Ref.~\cite{OneLoopPentagonandBox}.
As a cross-check we also evaluated the tensor integrals prior to applying the
projectors, following the methods of Refs.~\cite{FeynmanParameters} that
express every tensor integral in terms of Schwinger parameters.  These
integrals are in turn expressed in terms of scalar integrals with
shifted dimensions and higher powers of propagators. 
We use {\tt
  FIRE5}~\cite{FIRE} to reduce these integrals to elements of the standard basis
of scalar integrals.  The integrals are then shifted back to four
dimensions using dimension-shifting
formulas~\cite{DimensionShifting,FeynmanParameters}.  Both methods
yield identical results.

\begin{figure}[tb]
\captionsetup[subfigure]{labelformat=empty}
\begin{center}
\subfloat[(a)]{{\includegraphics[scale=0.6]{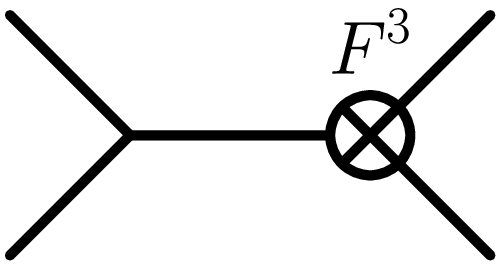}}}
\hspace{1cm}
\subfloat[(b)]{\includegraphics[scale=0.55]{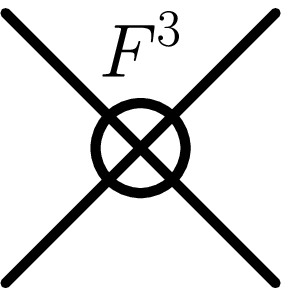}}
\end{center}
\vskip -.7cm
\caption[a]{\small Representative diagrams for (a) three- and (b)
  four-point $F^3$ insertions.}
 \label{f3ins}
 \end{figure}

We introduce linearized field strengths corresponding to each external
particle,
\begin{equation}
F_{i\,\mu\nu} \equiv k_{i\,\mu} \pol_{i\,\nu} - k_{i\,\nu} \pol_{i\,\mu}\,,
\label{FDef}
\end{equation}
in order to organize the results obtained from the projection
technique.  We express our results using Lorentz-invariant
combinations of these linearized field strengths.  For four-point
scattering in a parity-even theory, the only combinations at the
lowest mass dimension are~\cite{DoubleCopyCancel},
\begin{align}
(F_i F_j F_k F_l) &\equiv F_i^{\mu \nu}F_{j \,\nu \rho} F_k^{\rho \sigma} F_{l\, \sigma \mu}\,,
  \label{eq:F4} \\
(F_i F_j)(F_k F_l)&\equiv  F_i^{\mu\nu} F_{j \,\mu \nu} F_k^{\rho \sigma} F_{l\, \rho \sigma}\,.
\label{eq:F22}
\end{align}
These quantities are not symmetrized over the indices $i$, $j$, 
$k$, and $l$.

\def\Tr{\mathop{\rm Tr}\nolimits}
We need only one additional tensor for four-point scattering.  This
tensor can be expressed as a linear combination of terms of the form
$D^2F^4$.  It is, however, more convenient to express this tensor
as a matrix element with an insertion of an $F^3$ operator,
\begin{equation}
 F^3 \equiv \frac{1}{3} \Tr F^{\mu}{}_\nu F^{\nu}{}_\rho
 F^\rho{}_\mu\,,
\label{F3Def}
\end{equation}
where the trace is over color. The  gauge-invariant tensor is given by
\begin{equation}
T_{F^3} \equiv -i s t A^\tree_{F^3}(1,2,3,4)\,,
\label{TF3Def}
\end{equation}
using the four-point tree-level color-ordered amplitude with a single
insertion of the operator (\ref{F3Def}), as depicted in \fig{f3ins}.
As we see below, after applying the double-copy procedure, this
element of our basis is the one giving rise to the curvature-squared
matrix elements, as well as some of the anomalous ones.

\section{Result and Mapping to Supergravity}
\label{sec:maptogravity}

Using the tensors in Eqs.~(\ref{eq:F4}), (\ref{eq:F22}) and
(\ref{TF3Def}), we can write the supergravity amplitude as
follows\footnote{We write our results in the unphysical
  region where $s, t, u <0$; one can analytically continue to the
  physical region where $s>0$ and $t,u<0$ using $\ln(-s) \rightarrow
  \ln(s) - i \pi$.}\!\!,
\begin{align}
\begin{split}
    M^\oneloop_{\hmsg}(1,2,3,4) = \null & c_\Gamma 
       s t A^\tree_{\msym}(1,2,3,4)\\
   &  \times \Biggl[ \frac{t_8F^4}{stu} \biggl( -\frac{2}{\epsilon^2}\sum_{i< j}^3 
  s_{ij}\Bigl(\frac{-s_{ij}}{\mu^2}\Bigl)^{-\epsilon} + L_1(s,t,u) \biggr) \\
       &\null \hskip .8 cm  + \frac{T_{F^3}}{s t u} 
    + \biggl(\frac{4}{3} (F_1F_2F_3F_4) \biggl(\frac{1}{s t} + L_2(s,t,u) \biggr)\\ 
       & \null \hskip .8 cm +  (F_1F_2)(F_3F_4) \biggl(\frac{1}{s^2} + L_3(s,t,u) \biggr)
         + \text{cyclic(2,3,4)}  \biggr) \Biggr]\,,
    \end{split}
  \label{MatrixElement}
\end{align}
where $\mu$ is the usual scale parameter, $s_{12} = s, s_{23} = t, s_{13} = u$;
where
\begin{equation}
  c_\Gamma= \frac{\Gamma(1+\eps) \Gamma^2(1-\eps)}{(4 \pi)^{2-\eps}\Gamma(1-2\eps)}\,,
\end{equation}
is the usual one-loop prefactor,
  \begin{align}
  L_1(s,t,u) &=  - s\ln\Bigl(\frac{-s}{\mu^2}\Bigr)
                 - \frac{(2s^2+st+2t^2)}{2 u}\biggl(\ln ^2\Bigl(\frac{-s}{-t}\Bigr)+\pi^2\biggr) + \text{cyclic}(s,t,u)\,, \\
  L_2(s,t,u) &= \biggl[ - \frac{2s}{t^2u}\ln\Bigl(\frac{-s}{-u}\Bigr)
		 + \frac{1}{4 u^2}\biggl( \ln^2\Bigl(\frac{-s}{-t}\Bigr)+\pi^2\biggr)\nonumber\\
  &\hspace{15pt} + \frac{(s-2t)}{t^3}\biggl(\ln^2\Bigl(\frac{-s}{-u}\Bigr)+\pi^2\biggr) \biggr]
                 + (s\leftrightarrow t) \,,\\
  L_3(s,t,u) &=   \frac{1}{ s  t u}\biggl(-s\ln\Bigl(\frac{-s}{\mu^2}\Bigr) 
                 - t \ln\Bigl(\frac{-t}{\mu^2}\Bigr)
                 - u \ln\Bigl(\frac{-u}{\mu^2}\Bigr)\biggr) \nonumber \\ 
  &\hspace{15pt} + \frac{(t-u)}{s^3} \ln{\Bigl(\frac{-t}{-u}\Bigr)}
	         + \frac{(2s^2-tu)}{s^4} \biggl(\ln^2\Bigl(\frac{-t}{-u}\Bigr)+\pi^2\biggr)\,,
  \end{align}
and where we have used the combination 
\begin{equation}
t_8 F^4 = 2 (F_1F_2F_3F_4) - \frac{1}{2} (F_1F_2)(F_3F_4) + \text{cyclic}(2,3,4) \,,
\label{t8F4}
\end{equation}
familiar from the four-point one-loop type-I superstring amplitude.
The rank-$8$ tensor $t_8$ arises from the trace over the
fermionic zero-modes (see for instance\footnote{The
  $t_8$ tensor used here differs from the one in
  Ref.~\cite{Green1987sp} by an overall factor of $4$.}
Ref.~\cite{Green1987sp}).  The combination in \eqn{t8F4} is crossing
symmetric and is related to the Yang--Mills tree amplitude via
\begin{equation}
t_8 F^4 = -i s t A^{\tree}(1,2,3,4) =  -i s u A^{\tree}(1,2,4,3) =  -i t u A^{\tree}(1,3,2,4)\,.
\end{equation}
The amplitude in \eqn{MatrixElement} is ultraviolet-finite; the poles
in $\eps$ in \eqn{MatrixElement} are infrared ones.  

We have carried out a number of checks of the amplitude.  A simple
check is that the infrared singularity in \eqn{MatrixElement} matches
the known form~\cite{IRPapers},
\begin{equation}
\hskip -. cm 
M_{\hmsg}^{\text{1-loop}}\Bigl|_{\rm IR} = -
 M_{\hmsg}^{\text{tree}}
  \frac{2 c_\Gamma}{\epsilon^2} \sum_{i<j}^3 s_{ij} 
  \Bigl(\frac{-s_{ij}}{\mu^2}\Bigl)^{-\epsilon} \,.
\label{GravityIR}
\end{equation}
To see this we express the factors in front of the $1/\eps^2$ 
in \eqn{MatrixElement} in terms of the supergravity tree amplitude,
\begin{align}
s t A^\tree_\msym(1,2,3,4) \frac{t_8 F^4}{s t u} 
 = \null & -i s  A^\tree_\msym(1,2,3,4) A^{\tree}(1,2,4,3) =  M^\tree_{\hmsg}(1,2,3,4) \,,
\end{align}
where the last step uses the Kawai--Lewellen--Tye~(KLT)
relation~\cite{KLT} between tree-level gravity and Yang--Mills
amplitudes.  We have also compared the finite parts of all the
amplitudes with external scalars and gravitons to the results in
Ref.~\cite{OneLoopN4,Dunbar1995,CarrascoAnomaly} and found agreement.
The remaining fermionic amplitudes are related by supersymmetry Ward
identities.  We have checked that, prior to specializing to $D=4$, the
ultraviolet divergence cancels for $D<8$, as
expected~\cite{HalfMaxD5}.  In $D=8$, we match the prediction from the
heterotic string (see section 3.A.1 of Ref.~\cite{Tourkine2014ztw}) as
well as the calculation in Ref.~\cite{HalfMaxD5}.  It may also be
possible to compare our $D$-dimensional expression to the recent
$D=10$ prediction in Ref.~\cite{GreenArnab} obtained from
$M$-theory. However, performing this comparison would be nontrivial as
the divergences are quadratic in this dimension and hence depend on
the regulator. It would be interesting to study this connection
further.

The form in which we presented the amplitude in \eqn{MatrixElement}
makes the supersymmetry completely manifest, because it acts only on
the MSYM side of the double copy.  In addition, this form makes the
translation to gravity transparent.

We now show in some detail how this works for the case of external
gravitons.  In the double-copy construction, amplitudes with four
external gravitons can be built from integrands with purely gluonic
external states on both sides of the double copy.  As discussed in the
previous section, it is convenient to use linearized field strengths
in \eqns{eq:F4}{eq:F22} to write the answer. In order to translate to
gravity we do this on both sides of the double copy. From this
form, we can easily convert the linearized field strengths $F$ in our
formulas to a linearized Riemann tensor $R$ using the relation,
\begin{equation}
\frac{2}{\kappa} R_{i\,\mu\nu\rho\sigma} = F_{i\,\mu\nu} F_{i\,\rho\sigma} = 
(k_{i\,\mu} \pol_{i\,\nu} - k_{i\,\nu} \pol_{i\,\mu})\,
 ( k_{i\,\rho} \pol_{i\,\sigma} - k_{i\,\sigma} \pol_{i\,\rho})\,,
   \label{gravMapping}
\end{equation}
where the index $i$ refers to the particle label, just as in
\eqn{FDef}. In this equation the product of Yang--Mills polarization
vectors is identified as a graviton polarization tensor
via the replacement $\pol_{i\, \mu}
\pol_{i\, \nu} \rightarrow \pol_{i\, \mu\nu}$.  
The graviton is related to the metric via
$g_{\mu\nu}=\eta_{\mu\nu}+\kappa h_{\mu\nu}$, as in
Ref.~\cite{DoubleCopyCancel}. The factor of $2/\kappa$ is included in
\eqn{gravMapping} so that $R_{i\,\mu\nu\rho\sigma}$ is given by the
linearized Riemann tensor with the field $h_{\mu\nu}$ replaced by a
polarization tensor $\pol_{i\,\mu\nu}$. 

The contribution from the pure-gluon factor from MSYM is always a factor
of $stA^{\rm tree}= i t_8F^4$. Once we multiply the tensors from
both sides of the double-copy we then obtain the following combinations,
\begin{align}
  t_8 F^4 t_8 F^4 &\to t_8 t_8 R^4\,, \label{t8t8R4}  \\
  t_8 F^4(F_i F_j F_k F_l) &\to t_8(R_i R_j R_k R_l)\,, \label{t8R4} \\
  t_8 F^4(F_i F_j)(F_k F_l) &\to t_8 (R_i R_j)(R_k R_l)\,, \label{t8R22}
\end{align}
where
\begin{align}
(R_i R_j)^{\mu_1 \mu_2 \mu_3 \mu_4}(R_k R_l)^{\ \mu_5\mu_6\mu_7\mu_8}
 &\equiv R_i{}^{\mu_1 \mu_2 \nu \lambda} R_j{}^{\mu_3\mu_4}{}_{\nu\lambda}R_k{}^{\mu_5\mu_6 \rho \sigma}
       R_l{}^{\mu_7 \mu_8}{}_{\rho\sigma}\,,\\
(R_i R_j R_k R_l)^{\mu_1 \mu_2 \mu_3\mu_4 \mu_4 \mu_5 \mu_6 \mu_7 \mu_8}
&\equiv R_i{}^{\mu_1\mu_2 \nu \lambda} R_j{}^{\mu_3\mu_4}{}_{\lambda\rho}
        R_k{}^{\mu_5\mu_6 \rho \sigma} R_l{}^{\mu_7\mu_8}{}_{\sigma\nu} \,.
\end{align}
In ten dimensions \eqn{t8t8R4} is a component of the only
$\mathcal{N}=2$ superinvariant, whereas \eqns{t8R4}{t8R22}
are components of the two $\mathcal{N}=1$ superinvariants~\cite{SuperInvariants,
GreenArnab}.

The mapping of the final $T_{F^3}$ tensor to gravity may appear more
complicated than for the $F^4$-class tensors, because the former is
generated from a scattering amplitude with an $F^3$ insertion, as
previously illustrated in \fig{f3ins}.  A relatively simple way to obtain this
tensor is to use KLT relations for amplitudes extended to include
insertions of this higher-dimensional
operator~\cite{Broedel2012,HeZhang}.  This extension is in line with
expectations from string-theory KLT relations \cite{Kawai1986,
  BjerrumBohrHigherDim}, where the operator appears in the low-energy
effective action.  In Refs.~\cite{Broedel2012,HeZhang} it was
established that the KLT relations apply to $F^3$ operators as,
\begin{equation}
 s A^{\tree}(1,2,3,4) \times A^\tree_{F^3}(1,2,4,3) = i M^\tree_{R^2}(1,2,3,4)\,,
\label{f3KLT}
\end{equation}
where all particles are gluons on left-hand side of the equation, 
and all are gravitons on the right-hand side when the helicities of
each pair of gluons align.  Direct checks using
Feynman diagrams, starting from the Einstein action, confirm that the
Gauss--Bonnet insertion into a four-point gravity tree amplitude indeed
satisfies \eqn{f3KLT}~\cite{UnpublishedGB}.  Hence we see that the
tensor $T_{F^3}$ maps into the curvature-squared matrix elements in
gravity as follows,
\begin{align}
s t A^\tree(1,2,3,4) T_{F^3} = -i s u A^\tree(1,2,4,3) s t
 A^\tree_{F^3}(1,2,3,4) = s t u M^\tree_{R^2}(1,2,3,4) \,, 
\end{align}
where we used the crossing symmetry of $st A^\tree(1,2,3,4)$ and the
KLT relation in \eqn{f3KLT}.

After the complete map to linearized Riemann tensors, the graviton
amplitude takes the form,
\begin{align}
\begin{split}
M^\oneloop_{\hmsg} = \null & c_\Gamma
\Biggl[  M^\tree_{\hmsg} \biggl(-\frac{2}{\epsilon^2}\sum_{i<j}^3 s_{ij}
    \Bigl(\frac{-s_{ij}}{\mu^2}\Bigl)^{-\epsilon} + L_1(s,t,u)\biggr) \\
& \null \hskip .8 cm + {M}^\tree_{R^2} 
   + \bigg(\frac{4}{3}\,t_8(R_1R_2R_3R_4)\biggl(\frac{1}{st}+ L_2(s,t,u)\biggr)\\
& \null \hskip .8 cm + t_8(R_1R_2)(R_3R_4) \biggl(\frac{1}{s^2} + L_3(s,t,u)\biggr) 
+ \text{cyclic}(2,3,4)\bigg) \Biggr] \,.
\end{split}
\label{GravityResult}
\end{align}
The same construction works for any supergravity state. 
For all states in the supergravity multiplet, the same
pure Yang--Mills tensors feed into the corresponding supergravity
expressions; the differences are solely on the MSYM side of
the double copy.

It is remarkable that the coefficient of the curvature-squared matrix
element ${M}^\tree_{R^2}$ appearing in \eqn{GravityResult} is just a
simple number.  If the theory had a nonvanishing trace
anomaly~\cite{TraceAnomaly}, the coefficient of $M^\tree_{R^2}$
would have contained a $1/\eps$
divergence~\cite{OneLoopInfinity,GoroffSagnotti,GBPureGravity}.  In
our calculation the divergences are suppressed by an explicit factor
of $D-4 = 2 \eps$, (see, for example, Eq.~(2.11) of
Ref.~\cite{EnhancedIntegrals}) leaving a finite rational contribution.
From the perspective of the double copy, this $\eps/\eps$ effect also
generates the nonvanishing all-plus and single-minus one-loop
amplitudes associated with the $\text{U}(1)$ duality anomaly~\cite{CarrascoAnomaly}.  We
comment on this below.

\section{Curvature-Squared Multiplets and Divergences in Supergravity }
\label{sec:R2multiplets}

In the previous section we found curvature-squared contributions
to the effective action. In this section we describe these contribution
in more detail.

\subsection{Curvature-Squared Multiplets with Half-Maximal Supersymmetry}
In the full superamplitude, we find a term proportional to,
\begin{equation}
 s A^\tree_{\msym}(1,2,4,3) A^\tree_{F^3}(1,2,3,4)\,,
\label{fullKLTF3}
\end{equation}
which, as described in the previous section, contains the evanescent
matrix element of curvature operators.  In general
dimensions there exist several off-shell curvature-squared operators in gravity
theories. The two most important ones are the Gauss--Bonnet density
and the square of the Weyl tensor\footnote{There is another
  interesting curvature-squared operator, the Pontryagin density
  $^*R_{\mu\nu\rho\sigma}R^{\mu\nu\rho\sigma}$; but it is parity odd
  and hence it cannot appear in the amplitudes of parity-conserving
  theories.}\!\!, which respectively are given by,
\begin{align}
E_4 &= R_{\mu\nu\rho\sigma}R^{\mu\nu\rho\sigma}-4R_{\mu\nu}R^{\mu\nu}
      +R^2 \,,   \label{eq:gaussbonnet} \\
W^2 &= W_{\mu\nu\rho\sigma}\,W^{\mu\nu\rho\sigma}
= R_{\mu\nu\rho\sigma}R^{\mu\nu\rho\sigma}-2R_{\mu\nu}R^{\mu\nu}
     +\frac{1}{3}R^2 \,.
    \label{eq:weylsquare}
\end{align}
The difference between the two is,
\begin{equation}
  W^2 - E_4 = 2(R_{\mu\nu}R^{\mu\nu}-\frac{1}{3}R^2)\,,
\end{equation}
which vanishes on shell.  The single on-shell independent operator is
usually chosen to be the Gauss--Bonnet combination
(\ref{eq:gaussbonnet}). It is however a total derivative in four
dimensions, which implies that all curvature-squared matrix elements
are evanescent in this dimension~\cite{tHooftVeltman}.  A consequence
of this is the finiteness of pure-graviton amplitudes at one
loop~\cite{tHooftVeltman} in Einstein gravity, as these operators are
the only available counterterms.  (When matter is added to the
theory---even supersymmetric matter multiplets---generic divergences
do appear at one loop starting with amplitudes for four matter
particles~\cite{MatterDivergences}.)

Off-shell $R^2$ supermultiplets were constructed
long ago for $\mathcal{N}=1$ supergravity in four
dimensions~\cite{FerraraGB}, and more recently for $\mathcal{N}=2$
supergravity~\cite{ButterN2GB} using a version of $\NeqTwo$
superspace.  Very recently an $\NeqFour$ supersymmetric completion of
the Weyl-squared operator has been discussed in
Ref.~\cite{RenataAnomaly} in terms of linearized superfields in four
dimensions.  However, at the nonlinear
level no fully off-shell versions have been constructed to date
for any of the curvature-squared
multiplets.  This is unsurprising in light of the more general 
unsolved problem of constructing an off-shell $\NeqFour$ superspace.

\Eqn{fullKLTF3} also contains matrix elements related by supersymmetry
to the one corresponding to curvature-squared operators.  These must
arise from the $\NeqFour$ supersymmetric completion of the curvature
operators in \eqns{eq:gaussbonnet}{eq:weylsquare}. Therefore the existence of such
matrix element implies the existence of the corresponding $\NeqFour$
curvature-squared multiplets.  In particular, these matrix elements
should correspond to the single insertion of the operator discussed in
Ref.~\cite{RenataAnomaly} in four dimensions as all curvature-squared
operators are equivalent on shell.  However, we cannot analyze such
matrix elements strictly in four dimensions, because they will vanish
identically.

The double-copy construction provides additional information,
because it implies that completions of curvature-squared operators with
half-maximal supersymmetry should exist in any integer dimension $D\le10$ 
and that their on-shell matrix elements are given by the KLT product
of the $F^3$ operator insertion and ordinary MSYM amplitudes.   
The restriction to $D\le10$ arises because
that is the maximum dimension for a super-Yang--Mills theory.

The double-copy perspective also shows that an $\mathcal{N}\ge 5$
supersymmetric completion of curvature-squared
operators~\cite{deWitCurvatureSquare} cannot exist.  We have an
overall factor of $st A^\tree_{\msym}$ from the MSYM amplitude on the
one side of the double copy.  On the other side we would have an
$\mathcal{N}\geq 1$ super-Yang--Mills amplitude.  From the double-copy
perspective, in any dimension the $R^2$ terms correspond to an $F^3$
operator on this latter side.  We would then need a supersymmetric
completion of the $F^3$ operator, to make it compatible with $\NeqOne$
supersymmetry.  We know, however, that no such completion exists in four
dimensions because $F^3$ matrix element contributes only to all-plus
and single-minus helicity configurations; and these are forbidden by a
supersymmetric Ward identity~\cite{SWI}.  This also rules out
supersymmetric completions for these theories in any dimension $D>4$
because on shell there is only a single independent curvature-squared
invariant and one can choose the external momenta and states to live
in a four-dimensional subspace, and hence the same argument applies.

\subsection{Possible Effects at Higher Loops}

\begin{figure}[tb]
  \includegraphics[scale=0.5]{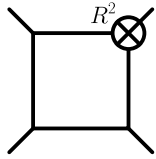}
\caption{Representative diagram for the insertion of the evanescent
  $R^2$ counterterm, affecting the two-loop divergence in pure-graviton
  amplitudes~\cite{GBPureGravity}.}
 \label{R2ins}
 \end{figure}

In the context of dimensional regularization, evanescent $R^2$
contributions such as the ones described here play a crucial role in
the two-loop divergences of pure
gravity~\cite{GoroffSagnotti,vandeVen}.  This happens because the
evanescent $R^2$ terms appear at one loop with a divergent coefficient
proportional to the trace anomaly.  While such terms do not contribute
in four dimensions, they do appear at two loops as subdivergences in the
dimensionally regulated amplitude, directly affecting the value of
the two-loop divergence~\cite{GBPureGravity}. One must then subtract a
one-loop $R^2$ counterterm insertion, as illustrated in \fig{R2ins}.  This
evanescent contribution becomes nonvanishing in dimensional
regularization where it modifies the two-loop divergence.  The net
result is a curious disconnect between the coefficient of the
dimensionally-regulated two-loop $R^3$ ultraviolet divergence of these
theories and the corresponding renormalization-scale dependence.  The
coefficient of the divergence depends on details of the
regularization, while the renormalization scale dependence is simple
and robust~\cite{GBPureGravity,GBPureGravitySimple}.

\begin{figure}[tb]
\captionsetup[subfigure]{labelformat=empty}
\begin{center}
\subfloat[ (a)]{{\includegraphics[scale=0.45]{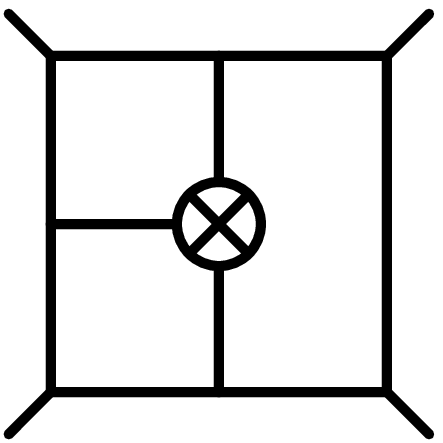}}}
\hspace{1cm}
\subfloat[(b)]{\includegraphics[scale=0.48]{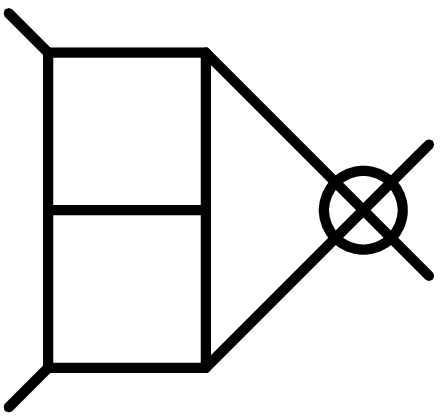}}
\end{center}
\vskip -.7cm
\caption[a]{Representative diagrams for insertions of the
  supersymmetric $R^2$ operator at three loops that could affect the
  four-loop divergence.}
 \label{threeloopins}
 \end{figure}

As shown in \eqn{GravityResult}, in $\NeqFour$ supergravity the $R^2$
contribution appears with a finite coefficient, so it cannot
contribute to possible two-loop divergences.  One may nonetheless
expect it to modify divergences at yet-higher loops.  Explicit
calculations reveal no divergences in $\NeqFour$ supergravity through
three loops~\cite{N4GravThreeLoops}, but unveil them at four
loops~\cite{FourLoopN4Sugra}.  The addition of supersymmetrization of
a curvature-squared operator as a local counterterm to the action is
not expected to have any physical consequences in the scattering
amplitudes, because it is evanescent.  The analysis in
Ref.~\cite{GBPureGravity} shows that it can however affect
divergences.  It would be interesting to study the effect of such
local counterterms on the known four-loop divergence calculated in
Ref.~\cite{FourLoopN4Sugra}. One may wonder whether such a finite
counterterm can be used to modify or even remove the four-loop
divergence.  The answer to this question would require a three-loop
computation with insertions of this operator, as illustrated in
\fig{threeloopins}.


\section{Evanescent Effects and the $\text{U}(1)$ Anomaly}
\label{sec:anomaly}

We now show that from the vantage point of the double copy that the
$\text{U}(1)$ anomalous contributions cannot be separated from the evanescent $R^2$
matrix elements, described in the previous section.  We first review
the anomaly and its manifestation in one-loop matrix
elements~\cite{CarrascoAnomaly}, before explaining how these effects 
are intertwined.

In order to describe the anomaly we recall some basic facts about the
spectrum of four-dimensional $\NeqFour$ supergravity and the
associated superamplitudes.  We focus here on pure $\NeqFour$ supergravity
with no matter multiplets.
The states of pure $\NeqFour$ supergravity fall into two
supermultiplets. One contains the positive-helicity graviton and its
superpartners~\cite{Cremmer}:
\begin{equation}
(h^{++}, \psi^{+}_a, A_{ab}^+,\chi^{+}_{abc},\phi^{-+}) \,,
  \label{eq:supermultiplets}
\end{equation}
where $h^{++}$ is the positive-helicity graviton, $\psi^{+}_a$ are the four
positive-helicity gravitinos, and so forth until the complex scalar
$\phi^{-+}$. The indices $a,b,c$ are $SU(4)$ $R$ symmetry indices. The
other supermultiplet is the CPT conjugate to the one above, containing
the negative-helicity graviton $h^{--}$ and the conjugate scalar
$\phi^{+-}$.  Seen through the lens of the double-copy, each multiplet
corresponds to the supermultiplet of MSYM multiplied by either a
positive- or negative-helicity gluon on the pure Yang--Mills side.
For instance the positive-helicity graviton arises from a
positive-helicity gluon on both sides of the double copy, and the
complex scalars come from negative-helicity gluons on one side and
positive-helicity gluons on the other side.

\begin{table}[tb]
\centering
\begin{tabular}{c c c@{\hskip 10pt} c}
\hline
\hline
Pure YM && $\NeqFour$ MSYM & \hskip .4 cm $\NeqFour$ Supergravity \\
\hline
$\langle g^- g^-g^+ g^+\rangle $ & $\otimes$ & $\langle g^- g^- g^+ g^+\rangle$ & 
          \hskip .4 cm $\langle h^{--}h^{--}h^{++}h^{++} \rangle$ \\
$\langle g^- g^+ g^+ g^+\rangle $ & $\otimes$ & $\langle g^- g^- g^+ g^+\rangle$ & 
          \hskip .4 cm $\langle h^{--}\phi^{-+}h^{++}h^{++} \rangle$ \\
$\langle g^+ g^+ g^+ g^+\rangle $ & $\otimes$ & $\langle g^- g^- g^+ g^+\rangle$ & 
         \hskip .4 cm $\langle \phi^{-+}\phi^{-+}h^{++}h^{++} \rangle$ \\
\hline
\end{tabular}
\caption{Top components of three of the five independent
  superamplitudes.  The other two are obtained from CPT conjugation.}
\label{tab:sectors}
\end{table}
Because not all the states of this theory are in a single
supermultiplet, the amplitudes are organized into different sectors
not directly related by supersymmetry. For each one of these sectors
there is an associated superamplitude.  A simple way to understand
this organization is via the double-copy construction.  The
supersymmetry Ward identities imply that the only nonvanishing
helicity amplitudes in MSYM are those in the
maximally-helicity-violating (MHV) sector corresponding to amplitudes
with two negative-helicity and two positive-helicity gluons $(g^- g^-
g^+ g^+)$ and their superpartners, which all sit in a single
superamplitude.  On the pure Yang--Mills side of the double copy, however,
there are three distinct types of amplitudes: all-plus $(g^+ g^+ g^+
g^+)$, single-minus $(g^- g^+ g^+ g^+)$, and two-minus or MHV
$(g^- g^- g^+ g^+)$, together with their parity conjugates. Hence there are
three distinct sectors of supergravity super-amplitudes, inherited from each of the
pure-Yang--Mills helicity configurations. In the all-plus and
single-minus pure Yang--Mills sectors the gluons do not have the same number of
negative or positive helicities as the gluons in the MSYM amplitude.
Because of this the corresponding $\NeqFour$ supergravity
superamplitudes do not contain four-graviton amplitudes, but have
mixed graviton--scalar amplitudes as their top components, as
illustrated in \tab{tab:sectors}. 

Ref.~\cite{MarcusAnomaly} showed that there exists an anomaly in an
abelian $\text{U}(1)$ subgroup of the $\text{SU}(1,1)$ duality group of $\NeqFour$
supergravity. This anomaly is manifested in the nonvanishing of
the amplitudes,
\begin{align}
M_{\hmsg} (1_{h^{--}},2_{\phi^{-+}},3_{h^{++}}, 4_{h^{++}}) & =   \frac{i}{(4 \pi)^2}  
            \frac{\spa1.2^2 \spa1.3^2 \spb 2.3^2 \spb3.4^4}{s t u} \,,
 \nonumber\\
M_{\hmsg} (1_{\phi^{-+}},2_{\phi^{-+}},3_{h^{++}},4_{h^{++}}) & =  
             \frac{i}{(4 \pi)^2} \spb3.4^4 \,, 
\label{eq:anomalousamp}
\end{align}
as well as those related by supersymmetry~\cite{CarrascoAnomaly}.  The spinor inner products
$\spa{a}.b$ and $\spb{a}.b$ follow the standard conventions in Ref.~\cite{ManganoParke}.
 The scalars carry a charge under the
$\text{U}(1)$ subgroup whereas the gravitons are uncharged and hence these
amplitudes violate conservation of this charge. At tree
level the charges are conserved because the amplitudes all vanish, but
at loop level they do not.  This anomaly can be traced back to
$\Ord(\eps)$ terms which interfere with a would-be $1/\eps$
divergence, leaving behind a rational term.  This is similar to the
way the chiral anomaly arises in dimensional
regularization~\cite{HVScheme}.

\begin{table}[tb]
  \centering
  \begin{tabular}{c@{\hskip 10pt} c}
  \hline
  \hline
   $\NeqFour$ Supergravity & 
          $\hskip .3 cm -i s A^{\tree}_{\msym}(1,2,4,3) A^{\tree}_{F^3}(1,2,3,4)$ \\
  \hline
   $\langle h^{--}h^{--}h^{++}h^{++} \rangle$ & $\hskip .3 cm 0$\\
   $\langle h^{--}\phi^{-+}h^{++}h^{++} \rangle$&  $-i \frac{\spa1.2^2 \spa1.3^2 \spb2.3^2}{s t u}\delta^{(8)}({\overline Q}) $\\
   $\langle \phi^{-+}\phi^{-+}h^{++}h^{++} \rangle$& 
             $\hskip .3 cm  2 i \, \delta^{(8)}({\overline Q}) $ \\
  \hline
  \end{tabular}
  \caption{Top components of the three independent
     sectors in four dimensions and corresponding superamplitudes. 
   }
  \label{tab:f3kltvalues}
\end{table}

As explained above, our calculation reveals evanescent contributions
in \eqn{fullKLTF3}, which are related to the supersymmetric completion
of the $R^2$ operator.  Mixed graviton--scalar amplitudes also receive
non-evanescent contributions from the same terms.  A simple way to see
this is by expressing the $F^3$ matrix element in a basis of
gauge-invariant tensors that has definite four-dimensional helicity
properties.  We give two such bases in the Appendix.  In the basis
with tensors that have definite crossing-symmetry properties, we find
that the $F^3$ matrix element is given by,
\begin{equation}
T_{F^3} = \frac{2 stu}{(s^2+t^2+u^2)}\, H^{(++++)} - H^{(-+++)}
 + \frac{2(s-t)(s-u)(t-u)}{3(s^2+t^2+u^2)^2} \, H^{\text{ev1}} 
- \frac{6 s t u}{(s^2+t^2+u^2)^2} \, H^{\text{ev2}}\,,
	\label{F3decomp}
\end{equation}
where the nonlocal denominators all
cancel to give a local expression for $T_{F^3}$.
This decomposition explicitly shows that $T_{F^3}$ has nonvanishing
contributions to the all-plus and single-minus helicity configurations,
with the rest of the tensor being evanescent in four dimensions.  This
gives some additional insight into the evanescent nature of the $R^2$ matrix
element in gravity.  The only nonvanishing amplitudes
on the MSYM side of the double copy have an MHV helicity configuration
$\hel --++$, whereas \eqn{F3decomp} shows that the $F^3$ matrix element
does not contribute to MHV amplitudes on the pure Yang--Mills side.  This
implies that the pure-graviton matrix elements vanish in four
dimensions.  More importantly, we see that this matrix element
contributes to the all-plus and single-minus helicities, thus
generating anomalous mixed graviton-scalar matrix elements after applying
the double-copy construction.

An alternative way to understand the different contributions of this
matrix element is to recall that in general dimension, a pair of
gluons is mapped via the double copy to a graviton, a dilaton and an
antisymmetric tensor. In four dimensions the antisymmetric tensor is
dual to a pseudoscalar that together with the dilaton combines into
the complex scalar discussed above.  The intertwining of the anomalous
and evanescent contributions in \eqn{f3KLT} therefore follows from the
entanglement of the graviton, dilaton and an antisymmetric tensor 
in the double-copy construction.

From the discussion above, we conclude that the $F^3$ KLT product in
\eqn{f3KLT} not only gives the evanescent curvature-squared matrix
elements, but it necessarily results in an anomalous contribution to
the amplitude.  It is striking that contributions to both can be
traced back to precisely the same term in the double copy.  The
anomalous contributions arising from $T_{F^3}$ are summarized in
\tab{tab:f3kltvalues}.  In this table the supermomentum delta function 
can be expanded as~\cite{ScatteringReview} 
\begin{equation}
\delta^{(8)}({\overline Q}) = 
  \delta^{(8)} \Bigl(\sum_{j=1}^4 \tilde\lambda^{\dot\alpha}_j \tilde\eta_{j a} \Bigr)
  = \prod_{a=1}^4 \sum_{i<j}^4 \spb{i}.{j} \tilde\eta_{i a} \tilde\eta_{j a},
\end{equation}
where we take the top component to be the one containing the factor
$\spb{3}.{4}^4$.  Comparing these to the anomalous amplitudes in
\eqn{eq:anomalousamp} we see that, while the amplitudes in the
single-scalar sector are fully contained in this term, those in the
two-scalar sector are off by an overall factor and receive additional
contributions that change the overall coefficient.

Finally, it is interesting to note that such anomalous and evanescent
effects will not appear in the one-loop amplitudes of $\mathcal{N} \ge 5$
supergravity.  The lack of anomalous one-loop amplitudes in $N\ge5$
supergravity has been recently explained from the vantage point of
super-invariants~\cite{RenataAnomaly}.  This, together with the
absence of evanescent effects, is understood in the double-copy
procedure as a consequence of the vanishing of the one-loop
all-plus and single-minus amplitudes in super-Yang--Mills theories.

\section{Conclusion}
\label{sec:conclusion}

In this paper we identified terms in the dimensionally regulated
one-loop four-point amplitude of pure $\NeqFour$ supergravity that can
be written as insertions of curvature-squared operators into matrix
elements.  Such terms are evanescent and vanish for four-dimensional
external states.  We also showed that these evanescent terms are
intertwined with contributions generated by the $\text{U}(1)$ duality
anomaly~\cite{MarcusAnomaly,CarrascoAnomaly}.  These two effects both
arise from rational pieces that result from an $\eps/\eps$
cancellation, where $\eps = (4-D)/2$ is the dimensional
regularization parameter. 

Both the anomaly and the evanescent curvature-squared terms may play a
central role in the ultraviolet properties of gravity theories.  As
explained in Ref.~\cite{CarrascoAnomaly} the anomaly in $\NeqFour$
supergravity gives contributions with a poor ultraviolet behavior.  We
also know that beyond one loop, evanescent effects contribute to
dimensionally regulated ultraviolet divergences in gravity
theories~\cite{GBPureGravity}.

We carried out our analysis using the double-copy
construction~\cite{BCJ,BCJLoop} of $\NeqFour$
supergravity~\cite{OneLoopN4} in terms of the corresponding pure
Yang--Mills and $\NeqFour$ MSYM amplitudes.  The double-copy
construction makes the on-shell supersymmetry manifest, because
$\NeqFour$ supergravity inherits the well-understood on-shell
superspace of MSYM theory.  By using formal polarization vectors on
the pure-Yang--Mills side of the double copy, we were able to evaluate
all one-loop four-point amplitudes of $\NeqFour$ supergravity
simultaneously. In the graviton sector we gave explicit conversion
formulas from gauge theory to gravity, using relations between
linearized Riemann tensors and Yang--Mills field strengths.  The
double-copy construction implies that completions of curvature-squared
operators with half-maximal supersymmetry should exist in any
dimension with $D\le10$ and that their on-shell matrix elements are
given by the KLT product of the $F^3$ operator insertion and ordinary
MSYM amplitudes.

There are a number of interesting avenues for future research.
Although it is is not known how to write the super-Gauss--Bonnet 
in an off-shell superspace, our paper provides all
components of four-point matrix elements of single insertions of these
operators. For the pure-graviton amplitude the Gauss--Bonnet operator
is the correct one for generating these matrix elements.  For amplitudes
with other external states, one would first need to systematically
write down a set of evanescent operators of the same dimension, feed
them through a tree-level matrix-element computation and then match
them to our evanescent matrix elements.  Once the combination of
operators leading to our evanescent matrix elements are found, one
can try to appropriately package the components into 
superfields.

We organized the one-loop amplitude in terms of gauge-invariant
tensors.  These and their associated projectors are described in the
appendix and given in the \textsl{Mathematica\/}
attachement~\cite{Ancillary}.  They are useful, not only for
$\NeqFour$ supergravity, but for any gauge-theory four-gluon amplitude
at any loop order.

In pure gravity the evanescent one-loop curvature-squared terms enter
with a coefficient proportional to $1/\eps$. Because of this, when
inserted as counterterms in a two-loop calculation they affect the
leading ultraviolet divergence~\cite{GBPureGravity}. In $\NeqFour$
supergravity these evanescent terms appear with a finite
coefficient. This means that they cannot affect divergences until
three loops or higher. Direct calculations show that the three-loop
divergences cancel~\cite{N4GravThreeLoops} and the first divergence
occurs at four loops~\cite{FourLoopN4Sugra}.  It is important to
understand the effect of evanescent and anomalous contribution on
higher-loop amplitudes, especially to see whether their contributions
can account for the four-loop divergence of $\NeqFour$ supergravity.
A direct study requires a three-loop computation.  An important step
in this direction would be to analyze the anomalous sector at two
loops in $\NeqFour$ supergravity and its relation to evanescent
effects.  In the longer term, understanding the role of anomalies and
evanescent effects more generally at higher loops appears to be
crucial in order to unravel the ultraviolet properties of supergravity
theories.

\section*{Acknowledgments}
We thank Rutger Boels, Thomas Dumitrescu, Michael Enciso, Michael Green, Henrik
Johansson, Radu Roiban, Arnab Rudra, Arkady Tseytlin and Mao Zeng for
many enlightening discussions. This work was supported in part by the
Department of Energy under Award Number DE-SC0009937 and in part by
the National Science Foundation under Grant Number NSF PHY-1125915.
J.P.-M. is supported by the U.S. Department of State through a
Fulbright Scholarship.  Part of DAK's work was carried out while a
Member of the Institute for Advanced Study.  DAK thanks the Institute
for its hospitality, and the Ambrose Monell Foundation for its support
of his stay there.


\newpage
\appendix
\section{Gauge-Invariant Tensors for Yang--Mills Four-Point Amplitudes}
\label{sec:appendix}

In this appendix, we describe two independent sets of Yang--Mills
kinematic tensors built out of physical polarization vectors $\pol_i$
and on-shell momenta $k_i$.  In both sets, the tensors are constrained
to be on-shell gauge invariant, that is vanishing under the
substitution $\pol_i\rightarrow k_i$ for each external leg
independently.  The tensors are polynomials in the dot products
$k_i\cdot \pol_j$, $\pol_i\cdot \pol_j$, and the Mandelstam invariants
$s$ and $t$.  They are thus free of poles by construction.  We also
organize the tensors to have definite symmetry properties under a
relevant symmetry, and to be diagonal in a four-dimensional helicity
basis.  The tensors are dimension-agnostic, and so the sets are not in
general diagonal in a basis of external states outside of four dimensions.
Both sets have seven tensors.

In the first set, each tensor represents kinematic parts of a color-ordered
amplitude, up to a function of $s$ and $t$.  Such amplitudes are 
invariant under a cyclic permutation of the external indices,
$i\rightarrow (i+1)\mod 4$, so we choose the tensors to have definite
symmetry properties under the cyclic shift.  An arbitrary function
can be split up into symmetric and antisymmetric combinations,
$f_{\pm}(s,t) = \frac12[f(s,t)\pm f(t,s)]$, so we choose the tensors
to be symmetric or antisymmetric.  It might seem simpler to choose
them to be symmetric; but for some of them, an antisymmetric form is
simpler.  In an amplitude, such antisymmetric tensors would then appear multiplied
by an antisymmetric function of $s$ and $t$.  We present this set in
the first subsection.

For the second set, each tensor represents one Yang--Mills copy in a
double-copy construction of an $\NeqFour$ supergravity amplitude,
where the other copy is given by the tree-level tensor.  These tensors
then suffice to construct the $\NeqFour$ supergravity four-point
amplitude at one and two loops.  These tensors are required to have
definite symmetry properties under the full permutation group acting
on the external indices.  We are interested only in the
one-dimensional representations of this group, so again each tensor
will either be completely invariant, or will change sign according to
the signature of a permutation.  We present this set in the second
subsection.

In the third subsection, we describe set of projection operators that can
be applied to an expression given in terms of polarization vectors and
momenta to obtain the (scalar) coefficients of the different basis tensors.

\subsection{Tensors with Definite Cyclic Symmetry}
\label{YMBasis}

We take the first element of the set of tensors with definite cyclic
properties to be the tensor of engineering dimension~4 that appears
in the tree amplitude,
\begin{equation}
\begin{aligned}
T^\tree=t_8F^4 & =
s\,(s+t)\,\eps_{1}\tcdot\eps_{4}\,\eps_{2}\tcdot\eps_{3}-s\,t\,\eps_{1}\tcdot\eps_{3}\,\eps_{2}\tcdot\eps_{4}
+t\,(s+t)\,\eps_{1}\tcdot\eps_{2}\,\eps_{3}\tcdot\eps_{4}
\\ &
-2\,(s+t)\,\eps_{1}\tcdot\eps_{4}\,k_{1}\tcdot\eps_{2}\,k_{1}\tcdot\eps_{3}
-2\,(s+t)\,\eps_{1}\tcdot\eps_{4}\,k_{1}\tcdot\eps_{2}\,k_{2}\tcdot\eps_{3}
\\ &
-2\,s\,\eps_{1}\tcdot\eps_{3}\,k_{1}\tcdot\eps_{2}\,k_{2}\tcdot\eps_{4}-2\,t\,\eps_{1}\tcdot\eps_{2}\,k_{1}\tcdot\eps_{3}\,k_{2}\tcdot\eps_{4}
-2\,(s+t)\,\eps_{1}\tcdot\eps_{2}\,k_{2}\tcdot\eps_{3}\,k_{2}\tcdot\eps_{4}
\\ &
-2\,t\,\eps_{2}\tcdot\eps_{4}\,k_{1}\tcdot\eps_{3}\,k_{3}\tcdot\eps_{1}-2\,t\,\eps_{2}\tcdot\eps_{4}\,k_{2}\tcdot\eps_{3}\,k_{3}\tcdot\eps_{1}
-2\,s\,\eps_{2}\tcdot\eps_{3}\,k_{2}\tcdot\eps_{4}\,k_{3}\tcdot\eps_{1}
\\ &
-2\,(s+t)\,\eps_{1}\tcdot\eps_{3}\,k_{1}\tcdot\eps_{2}\,k_{3}\tcdot\eps_{4}
-2\,(s+t)\,\eps_{1}\tcdot\eps_{2}\,k_{2}\tcdot\eps_{3}\,k_{3}\tcdot\eps_{4}
\\ &
-2\,(s+t)\,\eps_{2}\tcdot\eps_{3}\,k_{3}\tcdot\eps_{1}\,k_{3}\tcdot\eps_{4}
-2\,(s+t)\,\eps_{3}\tcdot\eps_{4}\,k_{1}\tcdot\eps_{2}\,k_{4}\tcdot\eps_{1}
\\ &
-2\,t\,\eps_{2}\tcdot\eps_{4}\,k_{1}\tcdot\eps_{3}\,k_{4}\tcdot\eps_{1}
-2\,(s+t)\,\eps_{2}\tcdot\eps_{4}\,k_{2}\tcdot\eps_{3}\,k_{4}\tcdot\eps_{1}
\\ &
-2\,(s+t)\,\eps_{2}\tcdot\eps_{3}\,k_{3}\tcdot\eps_{4}\,k_{4}\tcdot\eps_{1}
-2\,s\,\eps_{1}\tcdot\eps_{4}\,k_{1}\tcdot\eps_{3}\,k_{4}\tcdot\eps_{2}-2\,s\,\eps_{1}\tcdot\eps_{3}\,k_{2}\tcdot\eps_{4}\,k_{4}\tcdot\eps_{2}
\\ &
-2\,t\,\eps_{3}\tcdot\eps_{4}\,k_{3}\tcdot\eps_{1}\,k_{4}\tcdot\eps_{2}-2\,s\,\eps_{1}\tcdot\eps_{3}\,k_{3}\tcdot\eps_{4}\,k_{4}\tcdot\eps_{2}
-2\,(s+t)\,\eps_{3}\tcdot\eps_{4}\,k_{4}\tcdot\eps_{1}\,k_{4}\tcdot\eps_{2}
\,.
\end{aligned}
\label{TreeTensor}
\end{equation}
 It vanishes, of course, for the
$\hel ++++$ and $\hel -+++$ classes of helicities, and is nonvanishing
for MHV helicities $\hel --++$.  It is invariant
under cyclic shifts of the external legs.
We choose the remaining tensors to have
definite helicity properties as well.  We can give compact expressions
for the tensors in terms of the following combinations of the linearized
field-strength tensors defined in \eqn{FDef},
\begin{equation}
\begin{aligned}
\Ff{st}   & \equiv (F_1 F_2 F_3 F_4)\,, &
\Ff{tu}   & \equiv (F_1 F_4 F_2 F_3)\,, &
\Ff{us}   & \equiv (F_1 F_3 F_4 F_2)\,,\\
\Ft{s}    & \equiv (F_1 F_2)(F_3 F_4)\,, \hskip 5mm&
\Ft{t}    & \equiv (F_1 F_4)(F_2 F_3)\,, \hskip 5mm&
\Ft{u}    & \equiv (F_1 F_3)(F_4 F_2)\,,
\end{aligned}
\label{Fcombinations}
\end{equation}
along with the
$T_{F^3}$ tensor defined in \eqn{TF3Def}.  In terms
of these quantities, the basis tensors have the following expressions,
\begin{equation}
\begin{aligned}
T^\tree &=
-\frac{1}{2}(\Ft{s}+\Ft{t}+\Ft{u})
+2\,(\Ff{st}+\Ff{tu}+\Ff{us})
= t_8 F^4\,,\\
T^{\hel++++} &=
-2\,\Ff{st}+\frac{1}{2}\,(\Ft{s}+\Ft{t}+\Ft{u})
\,,\\
T^{\hel-+++} &=
-T_{F^3}-(\Ff{tu}-\Ff{us})\,(s-t)
+(\Ff{st}-\frac{1}{4}(\Ft{s}+\Ft{t}+\Ft{u}))\,(s+t)
\,,\\
T^{\hel--++} &=
\Ft{s}-\Ft{t}+2\,(\Ff{tu}-\Ff{us})
\,,\\
T^{\hel-+-+} &=
2\,\Ff{st}-\frac{1}{2}(\Ft{s}+\Ft{t}-\Ft{u})
\,,\\
T^{\ev1} &=
-(2\,\Ff{st}+\frac{3}{2}\,(\Ft{s}+\Ft{t}+\Ft{u}))\,(s+t)
+2\,(\Ff{us}\,(3\,s+t)+\Ff{tu}\,(s+3\,t))
\,\\
&=
-4\,(\Ff{tu}\,s+\Ff{us}\,t)
-(s+t)\,(8\,\Ff{st}-3\,T^\tree)
\,,\\
T^{\ev2} &=
-(2\,\Ff{st}-\frac{1}{2}(\Ft{s}+\Ft{t}+\Ft{u}))\,(s-t)
+2\,(\Ff{tu}-\Ff{us})\,(s+t)
\,\\
&=
4\,(\Ff{tu}\,s-\Ff{us}\,t)-(s-t)\,T^\tree
\,.
\end{aligned}
\label{SevenTensorsCyclic}
\end{equation}  

\begin{table}[tb]
\centering
\begin{ruledtabular}
\begin{tabular}{lcccc}
	Tensor &Dimension & Symmetry & Nonvanishing $D=4$ Helicity & $D=4$ Value \\
	\hline
	\multirow{2}{*}{$T^\tree$} &\multirow{2}{*}{4} &\multirow{2}{*}{$+$} &  $\hel--++$ &
        $%
        \spa{1}.{2}^{2}\,\spb{3}.{4}^{2}
        $ \\
	\cline{4-5}
	& & &  $\hel-+-+$ &
        $%
        \spa{1}.{3}^{2}\,\spb{2}.{4}^{2}
        $\\
	\hline
	$T^{\hel++++}$\, & 4 & $+$   & $\hel++++$ &
        $%
        \spb{1}.{3}^{2}\,\spb{2}.{4}^{2}
        $\\
	\hline
	$T^{\hel-+++}$ & 6 & $+$   & $\hel-+++$ &
        $%
        \spa{1}.{2}^{2}\,\spb{2}.{3}^{2}\,\spb{2}.{4}^{2}
        $\\
	\hline
	$T^{\hel--++}$ & 4 & $-$   & $\hel--++$ &
        $%
        \spa{1}.{2}^{2}\,\spb{3}.{4}^{2}
        $\\
	\hline
	$T^{\hel-+-+}$ & 4 & $+$   & $\hel-+-+$ &
        $%
        \spa{1}.{3}^{2}\,\spb{2}.{4}^{2}
        $\\
	\hline
	$T^{\ev1}$ & 6 & $+$   & --- & 0\\
	\hline
	$T^{\ev2}$ & 6 & $-$   & --- & 0\\
\end{tabular}
\end{ruledtabular}
\caption{Nonvanishing helicities and values for the color-ordered tensor basis. 
	Each tensor is also nonvanishing on the cyclic permutations and parity conjugates 
	of the indicated helicity states. The evanescent tensors vanish for all 
	four-dimensional helicities but are included in the table.}
\label{TensorHelicityTable} 
\end{table}

The first tensor is the tree-level tensor given above in \eqn{TreeTensor}.
The subsequent four tensors each are labeled by the class of four-dimensional
helicity configuration on which they are nonvanishing.
The final two tensors are nontrivial formal objects,
but vanish for all four-dimensional helicities. Outside of four dimensions,
they do not vanish, however, as demonstrated, for example, by the nonvanishing
value of the sum over states of each tensor multiplied by its conjugate.
They represent the kinematic part of evanescent operators in Yang--Mills theory.
In a slight abuse of language,  
we will therefore call them evanescent tensors.
Three other gauge-invariant tensors can be constructed, but these do not have
the correct symmetry properties to appear in color-ordered physical amplitudes.
The properties of all the tensors, as well as their values in four-dimensional 
helicity are summarized in \tab{TensorHelicityTable}.
The expressions for the tensors are also given in a companion
\textsl{Mathematica\/} file, \textsf{tensors-ym.m\/}.  
The notation there is,
\begin{equation}
\begin{aligned}
\textsf{ee[i,j]} &= \pol_i\cdot \pol_j\,, \hskip 10mm 
\textsf{ke[i,j]} &= k_i\cdot\pol_j\,, \hskip 10mm
\textsf{dot[i,j]} &= k_i\cdot k_j\,.
\end{aligned}
\label{Notation1}
\end{equation}
The seven tensors in \eqn{SevenTensorsCyclic} sequentially correspond to
\textsf{T[[i]]} in the file
for $i = 1, \ldots, 7$.
The spinor-valued expressions for the tensors
in four dimensions are also given that file,
with the seven values for each four-dimensional helicity configuration
recorded in \textsf{value[\textit{helicity-string}]}, for example
\textsf{value[``{+}{+}{+}{+}'']}.  These expressions employ the notation,
\begin{equation}
\textsf{spa[i,j]} = \spa{i}.j\,,\hskip 10mm
\textsf{spb[i,j]} = \spb{i}.j\,.
\label{Notation2}
\end{equation}

Conversely, we can express the linearized combinations~(\ref{Fcombinations})
in terms of the color-ordered tensors,
\begin{equation}
\begin{aligned}
\Ff{st} &=
-\frac{T^{\ev1}}{8\,(s+t)}-\frac{(s-t)\,T^{\ev2}}{8\,(s+t)^{2}}
+\frac{1}{4}\,T^\tree-\frac{s\,t\,T^{\hel++++}}{2\,(s+t)^{2}}
\,,\\
\Ff{tu} &=
\frac{T^{\ev2}}{4\,(s+t)}+\frac{1}{4}\,T^\tree
+\frac{t\,T^{\hel++++}}{2\,(s+t)}
\,,\\
\Ff{us} &=
-\frac{T^{\ev2}}{4\,(s+t)}+\frac{1}{4}\,T^\tree
+\frac{s\,T^{\hel++++}}{2\,(s+t)}
\,,\\
\Ft{s} &=
-\frac{T^{\ev1}}{4\,(s+t)}-\frac{(3\,s+t)\,T^{\ev2}}{4\,(s+t)^{2}}
+\frac{1}{2}\,T^\tree+\frac{1}{2}\,T^{\hel--++}-\frac{1}{2}\,T^{\hel-+-+} 
+\frac{s^{2}\,T^{\hel++++}}{(s+t)^{2}}
\,,\\
\Ft{t} &=
-\frac{T^{\ev1}}{4\,(s+t)}+\frac{(s+3\,t)\,T^{\ev2}}{4\,(s+t)^{2}}
+\frac{1}{2}\,T^\tree-\frac{1}{2}\,T^{\hel--++}-\frac{1}{2}\,T^{\hel-+-+}
+\frac{t^{2}\,T^{\hel++++}}{(s+t)^{2}}
\,,\\
\Ft{u} &=
T^{\hel-+-+}+T^{\hel++++}
\,,\\
T_{F^3} &=
-\frac{(s-t)\,T^{\ev2}}{2\,(s+t)}-T^{\hel-+++}
-\frac{2\,s\,t\,T^{\hel++++}}{s+t}
\,.
\end{aligned}
\label{SevenTensorsCyclicInverse}
\end{equation}

\subsection{Tensors with Definite Permutation Symmetry}
\label{preGBasis}

In this subsection, we present four-gluon kinematic tensors with
definite properties under the full permutation group.  These are
ultimately useful for decomposing $\NeqFour$ supergravity amplitudes
at one and two loops in a double-copy approach.  The tree
tensor~(\ref{TreeTensor}) is already fully crossing invariant, so we
take it to be the first tensor in this set as well, here calling
it $H^\tree$.
The remaining tensors are either invariant under all permutations of external labels,
or are multiplied by the signature of the permutation ($\pm1$).
We will call the latter signature-odd.  

\def\oddpoly{\alpha}
A signature-odd tensor will be multiplied by a signature-odd polynomial in $s$ and
$t$ in any physical amplitude.  Any invariant polynomial can also
appear as a tensor prefactor in an amplitude, of course.  All
invariant polynomials can be built out of products of two basic polynomials,
\begin{equation}
\begin{aligned}
\sigma_2(s,t,u) &= s^2 +t^2 +u^2 = 2(s^2 +s t +t^2)= -2 (st + tu + us)\,, \\
\sigma_3(s,t,u) &= s^3 +t^3 +u^3 = 3 stu\,,
\end{aligned}
\end{equation}
with a constant prefactor.
Any signature-odd polynomial is a product of an invariant
polynomial and the basic signature-odd
polynomial,
\begin{equation}
\oddpoly(s,t,u) = -(s-t)(t-u)(u-s) =  (s-t)(2s+t)(s+2t)\,.
\end{equation}
This polynomial satisfies the identity
\begin{equation}
2\,\oddpoly^2 = \sigma^3_2 - 6\,\sigma^2_3\,, 
\end{equation}
so that we need not consider powers of $\oddpoly$.

We can again express the tensors in terms of the linearized-field
strength quantities defined in \eqn{Fcombinations},
\begin{equation}
\begin{aligned}
H^\tree &=
-\frac{1}{2}(\Ft{s}+\Ft{t}+\Ft{u})
+2\,(\Ff{st}+\Ff{tu}+\Ff{us})
  = t_8 F^4\,,\\
H^{\hel++++} &=
\frac{3}{2}\,(\Ft{s}+\Ft{t}+\Ft{u})
-2\,(\Ff{st}+\Ff{tu}+\Ff{us})
\,,\\
H^{\hel-+++} &=
-T_{F^3}-\frac{4}{3}\,(\Ff{tu}\,s+\Ff{us}\,t-\Ff{st}\,(s+t))
\,,\\
H^{\mhv1} &=
-(\Ft{s}+2\,\Ff{tu})\,s-(\Ft{t}+2\,\Ff{us})\,t
+(2\,\Ff{st}+\Ft{u})\,(s+t)
\,,\\
H^{\mhv2} &=
\Ft{u}\,(s-t)\,(s+t)+\Ft{t}\,t\,(2\,s+t)
-\Ft{s}\,s\,(s+2\,t)
\,,\\
H^{\ev1} &=
4\,(\Ff{st}\,(s-t)\,(s+t)+\Ff{us}\,t\,(2\,s+t)
-\Ff{tu}\,s\,(s+2\,t))
\,,\\
H^{\ev2} &=
(\Ft{s}+\Ft{t}+\Ft{u})\,(s^{2}+s\,t+t^{2})
-4\,(\Ff{tu}\,t\,(s+t)-s\,(\Ff{st}\,t-\Ff{us}\,(s+t)))
\,.\\
\end{aligned}
\label{SevenTensorsCrossing}
\end{equation}  

The second and third tensors are again labeled by the four-dimensional
helicity class for which they are nonvanishing; the fourth and fifth 
are both nonvanishing for all MHV helicities.  The last two are again
``evanescent'', in the sense that they are nonvanishing outside of
four dimensions but vanish for all four-dimensional helicity configurations.
(As in \sect{YMBasis}, they do not include factors of $1/\eps$ that would
be needed to yield a nonvanishing result in four dimensions.)

\begin{table}[tbh]
\centering
\begin{ruledtabular}
	\begin{tabular}{lcccc}
	Tensor & Dimension  & Signature & Nonvanishing $D=4$ Helicity & $D=4$ Value \\
	\hline
	$H^\tree$ & 4 & even & $\hel--++$ &
        $%
        \spa{1}.{2}^{2}\,\spb{3}.{4}^{2}
        $ \\
	\hline
	$H^{\hel++++}$\, & 4 & even & $\hel++++$ &
        $%
        \spb{1}.{4}^{2}\,\spb{2}.{3}^{2}+\spb{1}.{3}^{2}\,\spb{2}.{4}^{2}
        +\spb{1}.{2}^{2}\,\spb{3}.{4}^{2}
        $\\
	\hline
	$H^{\hel-+++}$ & 6 & even & $\hel-+++$ &
        $%
        \spa{1}.{2}^{2}\,\spb{2}.{3}^{2}\,\spb{2}.{4}^{2}
        $\\
	\hline
	$H^{\mhv1}$ & 6 & even & $\hel--++$ &
        $%
        \spa{1}.{2}^{3}\,\spb{1}.{2}\,\spb{3}.{4}^{2}
        $\\
	\hline
	$H^{\mhv2}$ & 8  & odd & $\hel--++$ &
        $%
        (s+2\,t)\,\spa{1}.{2}^{3}\,\spb{1}.{2}\,\spb{3}.{4}^{2}
        $\\
	\hline
	$H^{\ev1}$ & 8  & odd & --- & 0\\
	\hline
	$H^{\ev2}$ & 8  & even & --- & 0\\
\end{tabular}
\end{ruledtabular}
\caption{Nonvanishing helicities and values for the pregravity tensor basis.  
	Each tensor is also nonvanishing on the permutations and parity
	conjugates of the indicated helicity states.  The evanescent tensors
	vanish for all four-dimensional helicities but are included in the table.} 
	\label{N=4TensorHelicityTable}
\end{table}

The expressions for the tensors are also given in a companion
\textsl{Mathematica\/} file, \textsf{tensors-neq4gr.m\/},
with \textsf{H[[i]]}, $i=1,\ldots,7$ corresponding in order
to the tensors in \eqn{SevenTensorsCrossing}. 
The spinor-valued expressions for the tensors in four dimensions are 
also given in that file; as in \sect{YMBasis}, the seven values
for each four-dimensional helicity configuration given by 
\textsf{value[{\it helicity-string\/}]}.  The notation follows
\eqns{Notation1}{Notation2}.
The properties of the tensors are summarized in \tab{N=4TensorHelicityTable}.

Because these tensor have definite properties under permutations, we 
can connect them straightforwardly to matrix elements of corresponding operators
after the double copy.  A few examples would be,
\begin{align}
\begin{split}
\sigma_2 t_8F^4t_8F^4  &\leftrightarrow  t_8t_8D^4R^4  \,, \\
t_8F^4(u \Ff{st} + s \Ff{tu} + t \Ff{us})  &\leftrightarrow  t_8 \text{tr}(D^2R^4)  \,,  \\
\sigma_2 t_8F^4 (\Ft{s} + \Ft{u} + \Ft{t}) &\leftrightarrow  t_8 (\text{tr}(DR)^2)^2 \,.
\end{split}
\end{align}

\subsection{Projectors for Basis Tensors}

In this subsection, we present a set of projectors that can be
used to obtain the scalar coefficients of the basis tensors
for an expression given in terms of polarization vectors
and momenta.  When applied to an integrated expression for
an amplitude, the resulting decomposition will reproduce the
original expression; when applied to an integrand, there may
be a total-derivative discrepancy that
will integrate to zero.

We define an inner product $\statesum$ of
a polarization vector and its conjugate to be given by the sum over
states,
\begin{equation}
\eps_i^{*\mu}\statesum\eps_i^\nu = 
\sum_{{\rm states}\, h} \eps^{*(h),\mu}_i \eps^{(h),\nu}_i
= -g^{\mu\nu} + \frac{k_i^{\mu}q_{\phantom{i}}^{\nu} + q_{\phantom{i}}^{\mu}k_i^{\nu} }{q\cdot k_i}\,,
\end{equation}
where $q$ is a null reference vector not collinear to any external
momentum.  (It is similar to a lightcone-gauge vector.)
In four dimensions, the state sum becomes,
\begin{equation}
\sum_{{\rm states}\, h} \eps^{*(h),\mu}_i \eps^{(h),\nu}_i
=\sum_{h=\pm} \eps^{*(h),\mu}_i \eps^{(h),\nu}_i
=\sum_{h=\pm} \eps^{(-h),\mu}_i \eps^{(h),\nu}_i\,,
\end{equation}
where the sum is over vector helicities.

In all dimensions, the projector onto the $j$th tensor is then given by,
\begin{equation}
P_j = c^{\phantom{*}}_{ji} T_i^*\,,
\end{equation}
where the matrix $c$ is the inverse of the (symmetric)
inner product matrix $m$,
whose elements are given by,
\begin{equation}
m_{ij} = T_i^*\statesum T_j\,.
\end{equation}
The coefficient of $T_j$ in an expression $X$ is given by $P_j\statesum X$.

Each basis has a corresponding set of projectors; the projectors
for the cyclicly-organized basis described in \sect{YMBasis} are
given alongside the tensors and helicity
values in \textsf{tensors-ym.m}, where the projector $P_j$
onto $T_j$ is given by \textsf{P[[j]]}.  The expressions make use
of the following notation in addition to that in \eqn{Notation1},
\begin{equation}
\textsf{cc[i,j]} = \pol^*_i\cdot \pol^*_j\,,\hskip 10mm
\textsf{kc[i,j]} = k_i\cdot\pol^*_j\,,\hskip 10mm
\textsf{chi} = t/s\,,\hskip 10mm
\textsf{d\/} = D\,.
\end{equation}
In four dimensions, $m$ has rank 5, as expected from the nature of
$T_5$ and $T_6$.  In six dimensions, it has rank 7, showing indirectly
that there are some helicities with non-vanishing values for these two
tensors.  The corresponding projectors for the basis of
\sect{preGBasis} organized under the full crossing symmetry are given
in \textsf{tensors-neq4gr.m}.  The projector matrix again has rank~5
in four dimensions, and rank~7 in six dimensions.


\bibliographystyle{apsrev}


\end{document}